\documentclass[11pt,english]{article}
\usepackage[latin9]{inputenc}
\usepackage{mathtools}
\usepackage{amsmath}
\usepackage{amssymb}

\makeatletter
\usepackage{amsmath}
\usepackage{amsthm}
\usepackage{amssymb}

\theoremstyle{plain}    
\newtheorem{theorem}{Theorem}[section] 
 
\theoremstyle{plain}    
\newtheorem{theorem-seq}{Theorem}

\newenvironment{myproof}[1][Proof.]{\par
\pushQED{\qed}%
\normalfont \topsep6\p@\@plus6\p@\relax
\trivlist
\item[\hskip\labelsep
\bfseries
#1]\ignorespaces
}{%
\popQED\endtrivlist\@endpefalse
}

\theoremstyle{plain}
\newtheorem*{rep@theorem}{\rep@title}

\newcommand{\newreptheorem}[2]{%
\newenvironment{rep#1}[1]{%
 \def\rep@title{#2 ##1}%
 \begin{rep@theorem}}%
 {\end{rep@theorem}}
}


\newcommand{\eqdef}{\stackrel{\rm{def}}{=}}
\DeclareMathOperator{\poly}{poly}

\newcommand{\E}{\mathbb{E}}

\newcommand{\N}{\mathbb{N}}

\newcommand{\R}{\mathbb{R}}

\newcommand{\F}{\mathbb{F}}

\newcommand{\dist}{\mathrm{dist}}

\renewcommand{\P}{\mathbf{P}}

 \theoremstyle{plain}
 \theoremstyle{definition}
 \newtheorem{remark}[theorem]{Remark}
 \theoremstyle{plain}    
 \newtheorem{fact}[theorem]{Fact} 
 \theoremstyle{definition}
 \newtheorem{definition}[theorem]{Definition}
 \theoremstyle{plain}    
 \newtheorem{lemma}[theorem]{Lemma}  
 \newreptheorem{lemma}{Lemma}
 \theoremstyle{plain}    
 \newtheorem{proposition}[theorem]{Proposition} 
 \theoremstyle{plain}    
 \newtheorem{claim}[theorem]{Claim}
 \theoremstyle{definition}    
 \newtheorem*{acknowledgement*}{Acknowledgement} 

\usepackage{xcolor}
\usepackage{mathtools} 
\usepackage{fullpage}
\usepackage{relsize}

\def\notes{1}
 \newcommand{\knote}[1]{\ifnum\notes=1{{\sf\color{blue} [Swastik's comment: #1]}}\fi}
 \newcommand{\nnote}[1]{\ifnum\notes=1{{\sf\color{blue} [Noga's comment: #1]}}\fi}
 \newcommand{\ssnote}[1]{\ifnum\notes=1{{\sf\color{blue} [Shubhangi's comment: #1]}}\fi}
 \newcommand{\onote}[1]{\ifnum\notes=1{{\sf\color{blue} [Or's comment: #1]}}\fi}

\newcommand{\set}[1]{\left\{{#1}\right\}}

\makeatother

\usepackage{babel}
\begin{document}

\title{High rate locally-correctable and locally-testable codes\\
with sub-polynomial query complexity%
\thanks{A preliminary version of this work appeared as~\cite{Meir-ECCC}.%
}}

\author{Swastik Kopparty%
\thanks{Department of Mathematics \& Department of Computer Science, Rutgers
University, Piscataway NJ 08854, USA. Supported in part by a Sloan
Fellowship and NSF grant CCF-1253886. \texttt{swastik.kopparty@gmail.com}%
}\and Or Meir%
\thanks{Department of Computer Science and Applied Mathematics, Weizmann Institute
of Science, Rehovot 76100, Israel. This
research was carried out when Meir was supported in part by the Israel
Science Foundation (grant No. 460/05). \texttt{or.meir@weizmann.ac.il}%
}\and Noga Ron-Zewi%
\thanks{ School of Mathematics, Institute for Advanced Study, Princeton, NJ,
USA. Supported in part by the Rothschild fellowship
and NSF grant CCF-1412958. \texttt{nogazewi@ias.edu} %
}\and Shubhangi Saraf%
\thanks{Department of Mathematics \& Department of Computer Science, Rutgers
University, Piscataway NJ 08854, USA. Supported in part by NSF grant
CCF-1350572. \texttt{shubhangi.saraf@gmail.com}%
} }
\maketitle
\begin{abstract}
In this work, we construct the first locally-correctable
codes (LCCs), and locally-testable codes (LTCs) with constant rate,
constant relative distance, and sub-polynomial query complexity. Specifically,
we show that there exist binary LCCs and LTCs with block length $n$, constant
rate (which can even be taken arbitrarily close to 1), constant relative distance,
and query complexity $\exp(\tilde{O}(\sqrt{\log n}))$. Previously
such codes were known to exist only with $\Omega(n^{\beta})$ query complexity (for constant $\beta > 0$),
and there were several, quite different, constructions known.

Our codes are based on a general distance-amplification method of
Alon and Luby~\cite{AL96_codes}. We show that this method interacts
well with local correctors and testers, and obtain our main results
by applying it to suitably constructed LCCs and LTCs in the non-standard
regime of \emph{sub-constant relative distance}.

Along the way, we also construct LCCs and LTCs over large alphabets,
with the same query complexity $\exp(\tilde{O}(\sqrt{\log n}))$, which additionally have the property
of approaching the Singleton bound: they have almost the best-possible
relationship between their rate and distance. This has the surprising
consequence that asking for a large alphabet error-correcting code
to further be an LCC or LTC with $\exp(\tilde{O}(\sqrt{\log n}))$
query complexity does not require any sacrifice in terms of rate and
distance! Such a result was previously not known for any $o(n)$ query
complexity.

Our results on LCCs also immediately give locally-decodable codes
(LDCs) with the same parameters.
\end{abstract}
\newpage{}

\section{Introduction}

\global\long\def\sW{\mathsmaller{W}}
\global\long\def\tW{\tau_{\sW}}
\global\long\def\rW{r_{\sW}}
\global\long\def\dW{\delta_{\sW}}
\global\long\def\nW{n_{\sW}}
\global\long\def\iW{i_{\sW}}
\global\long\def\AW{A_{\sW}}
\global\long\def\P{\mathcal{P}}
\global\long\def\RS{RS}
\global\long\def\H{\mathbb{H}}
\global\long\def\E{\mathcal{E}}
Locally-correctable
codes~\cite{BFLS91,STV01,KT00} and locally-testable codes~\cite{FS95,RS96,GS02}
are codes that admit local algorithms for decoding and testing respectively.
More specifically: 
\begin{itemize}
\item We say that a code $C$ is a \textsf{locally-correctable code (LCC)}%
\footnote{There is a closely related notion of locally decodable codes (LDCs)
that is more popular and very well studied. All our results for LCCs
hold for LDCs as well, see discussion at the end of the introduction.%
} if there is a randomized algorithm that, when given a string $z$
that is close to a codeword $c\in C$, and a coordinate $i$, computes~$c_{i}$
while making only a small number of queries to~$z$. 
\item We say that a code $C$ is a \textsf{locally-testable code (LTC)}
if there is a randomized algorithm that, when given a string $z$,
decides whether $z$ is a codeword of~$C$, or far from $C$, while
making only a small number of queries to~$z$. 
\end{itemize}
The number of queries that are used by the latter algorithms is called
the \textsf{query complexity}.

Besides being interesting in their own right, LCCs and LTCs have also
played important roles in different areas of complexity theory, such
as hardness amplification and derandomization (see e.g. \cite{STV01}),
and probabilistically checkable proofs \cite{AS98,ALMSS98}. It is
therefore a natural and well-known question to determine what are
the best parameters that LCCs and LTCs can achieve.

LCCs and LTCs were originally studied in the setting where the query
complexity was either constant or poly-logarithmic. In those settings,
it is believed that LCCs and LTCs must be very redundant, since every
bit of the codeword must contain, in some sense, information about
every other bit of the codeword. Hence, we do not expect such codes
to achieve a high rate. In particular, in the setting of constant
query complexity, it is known that linear LCCs cannot have constant
rate~\cite{KT00,WW05,W07}%
\footnote{\cite{KT00,WW05,W07} proved a lower bound for the related notion
of LDCs. Since every linear LCC is also an LDC, their lower bound
applies to linear LCCs as well.%
}, and that LTCs with certain restrictions cannot have constant rate~\cite{DK11,BV12}.
On the other hand, the best-known constant-query LCCs have exponential
length%
\footnote{For example, a constant-degree Reed-Muller code is such an LCC.%
}, and the best-known constant-query LTCs have quasi-linear length
(see e.g.~\cite{BS08,D07,V15}).

It turns out that the picture is completely different when allowing
the query complexity to be much larger. In this setting, it has long
been known that one can have LCCs and LTCs with constant rate and
query complexity $O(n^{\beta})$ for constant $\beta>0$ \cite{BFLS91,RS96}.
More recently, it has been discovered that both LCCs~\cite{KSY14,GKS13,HOW13}
and LTCs~\cite{V11_tensor_robustness,GKS13} can simultaneously achieve
rates that are arbitrarily close to~$1$ and query complexity $O(n^{\beta})$
for an arbitrary constant $\beta>0$. This is in contrast with the
general belief that local correctability and testability require much
redundancy.

In this work, we show that there are LCCs and LTCs with constant rate
(which can in fact be taken to be arbitrarily close to 1) and constant
relative distance,
whose associated local algorithms have $n^{o(1)}$ query complexity
and running time. We find it quite surprising in light of the fact
that there were several quite different constructions of LCCs and
LTCs~\cite{BFLS91,RS96,KSY14,V11_tensor_robustness,GKS13,HOW13}
with constant rate and constant relative distance, all of which had
$\Omega(n^{\beta})$ query complexity. 

Furthermore, we show that over large alphabets, such codes can approach
the Singleton bound: they achieve a tradeoff between rate and
distance which is essentially as good as possible for general error-correcting
codes. Such a result was previously not known for any $o(n)$ query
complexity. This means that, remarkably, local correctability and
local testability with $n^{o(1)}$ queries over large alphabets is
not only possible with constant rate and constant relative distance,
but it also does not require ``paying'' anything in terms of rate
and relative distance.

We first state our theorems for the binary alphabet. 
\begin{theorem}
[\label{thm:binary-LCCs}Binary LCCs with sub-polynomial query complexity]
For every $r\in(0,1)$, there exist $\delta>0$ and an explicit infinite
family of binary linear codes $\left\{ C_{n}\right\} _{n}$ satisfying: 
\begin{enumerate}
\item $C_{n}$ has block length $n$, rate at least $r$, and relative distance
at least $\delta$, 
\item $C_{n}$ is locally correctable from $\frac{\delta}{2}$-fraction
of errors with query complexity and running time at most $\exp(\sqrt{\log n\cdot\log\log n})$. 
\end{enumerate}
\end{theorem}

\begin{theorem}
[\label{thm:binary-LTCs}Binary LTCs with sub-polynomial query complexity]
For every $r\in(0,1)$, there exist $\delta>0$ and an explicit infinite
family of binary linear codes $\left\{ C_{n}\right\} _{n}$ satisfying: 
\begin{enumerate}
\item $C_{n}$ has block length $n$, rate at least $r$, and relative distance
at least $\delta$, 
\item $C_{n}$ is locally testable with query complexity and running time
at most $\exp(\sqrt{\log n\cdot\log\log n})$. 
\end{enumerate}
\end{theorem}
The binary LCCs and LTCs in the above theorems are obtained by first
constructing LCCs and LTCs over large alphabets, and then concatenating
them with binary codes. The following theorems describe these large
alphabet LCCs and LTCs, which in addition to having sub-polynomial
query complexity, also approach the Singleton bound. 
\begin{theorem}
[\label{main-thm-LCCs}LCCs with sub-polynomial query complexity
approaching the Singleton bound]For every $r\in(0,1)$, there exists
an explicit infinite family of linear codes $\left\{ C_{n}\right\} _{n}$
satisfying: 
\begin{enumerate}
\item $C_{n}$ has block length $n$, rate at least $r$, and relative distance
at least $1-r-o(1)$, 
\item $C_{n}$ is locally correctable from $\frac{1-r-o(1)}{2}$-fraction
of errors with query complexity and running time at most $\exp(\sqrt{\log n\cdot\log\log n})$,
\item The alphabet of $C_{n}$ is of size at most $\exp(\exp(\sqrt{\log n\cdot\log\log n}))$.
\end{enumerate}
\end{theorem}

\begin{theorem}
[\label{main-thm-LTCs}LTCs with sub-polynomial query complexity
approaching the Singleton bound]For every $r\in(0,1)$, there exists
an explicit infinite family of linear codes $\set{C_{n}}_n$ satisfying: 
\begin{enumerate}
\item $C_{n}$ has block length n, rate at least $r$, and relative distance
at least $1-r-o(1)$, 
\item \emph{$C_{n}$} is locally testable with query complexity and running
time at most $\exp(\sqrt{\log n\cdot\log\log n})$,
\item The alphabet of $C_{n}$ is of size at most $\exp(\exp(\sqrt{\log n\cdot\log\log n}))$.
\end{enumerate}
\end{theorem}
The above theorems are proved in Sections~\ref{sec:LCCs} and~\ref{sec:LTCs}.
\begin{remark}
If we were only interested in LCCs and LTCs with $O(n^{\beta})$ query
complexity (for arbitrary $\beta$), we could have constructed binary
codes that meet the \emph{Zyablov bound}, which is the best-known rate-distance
tradeoff for explicit binary codes. Furthermore, we could have constructed
codes over constant-size alphabet that approach the Singleton bound
(rather than having alphabet of super-constant size).

Moreover, our results imply the existence of \emph{non-explicit} binary LCCs/LTCs with query complexity $\exp(\sqrt{\log n\cdot\log\log n})$ that meet the Zyablov bound. This follows by concatenating the codes of Theorems \ref{main-thm-LCCs} and~\ref{main-thm-LTCs} with (non-explicit) Gilbert-Varshamov codes \cite{G52,V57}.
\end{remark}

\paragraph*{The Alon-Luby distance-amplification.}

Our constructions are based on the distance-amplification technique
of~\cite{AL96_codes}. This distance amplifier, based on a $d$-regular
expander, converts an error-correcting code with relative distance
$\gg1/d$ into an error-correcting code with larger relative distance
$\delta$, while reducing the rate only by a factor of $\approx(1-\delta)$.
Thus for a large enough constant $d$, if we start with a code of
rate $1-\varepsilon$ and relative distance $\gg1/d$, where $\varepsilon\ll\delta$,
then after distance amplification with a $d$-regular expander, we
get a code with rate $(1-\delta)(1-\varepsilon)\approx(1-\delta)$
and relative distance $\delta$. 

The original application of this technique in~\cite{AL96_codes}
was to construct linear-time erasure-decodable codes approaching the
Singleton bound. In addition to the above distance-amplification technique,
\cite{AL96_codes}~constructed a linear-time erasure-decodable code (not approaching the Singleton bound)
which could be used as the input code to the amplifier. The main result
of~\cite{AL96_codes} then follows from the fact that distance amplification
via a \emph{constant-degree} expander preserves linear-time erasure-decodability.

Subsequent applications of this distance-amplification technique followed
a similar outline. One first constructs codes with high rate with
some (possibly very small) constant relative distance and a certain
desirable property. Then, applying distance amplification with a (possibly
very large) constant-degree expander, one obtains a code with a much
better tradeoff between its rate and relative distance. Finally one
shows that the distance amplification with a constant degree
expander preserves the desirable property. This scheme was implemented
in \cite{GI05}, who constructed codes that can be decoded in linear
time from \emph{errors} (rather than \emph{erasures}), and in~\cite{GI02,GR08_folded_RS},
who constructed capacity-achieving list-decodable codes with constant
alphabet.

\paragraph*{Our observations.}

The first main observation of this paper is that the distance-amplification
technique also preserves the property of being an LCC or an LTC. Specifically,
if we start with an LCC or LTC with query complexity $q$, and then
apply distance amplification with a $d$-regular expander, then the
resulting code is an LCC/LTC with query complexity $q \cdot \poly(d)$.

The next main observation is that this connection continues to hold
even if we take $d$ to be super-constant, and take the LCC or LTC
to have sub-constant relative distance $\Theta(1/d)$ (and then we
only require the LCC to be able to correct strings whose distance
from the code is within some constant fraction of the minimum distance
of the code). This is potentially useful, since we only blow up the
query complexity by a factor of $\poly(d)$, and perhaps LCCs/LTCs
with high rate and sub-constant relative distance can have improved
query complexity over their constant relative distance counterparts.

Finally, we show that existing families of high rate LCCs and LTCs
can achieve sub-polynomial query complexity if we only require them
to have sub-constant relative distance. Specifically, multiplicity
codes~\cite{KSY14} in a super-constant number of variables give
us the desired LCCs, and super-constant-wise tensor products~\cite{V11_tensor_robustness}
give us the desired LTCs. 

As far as we are aware, there have been no previous uses of this distance-amplification
technique using an expander of super-constant degree.

More generally, we wish to draw attention to the technique of~\cite{AL96_codes}.
We believe that it should be viewed as a general scheme for improving
the rate-distance tradeoff for codes with certain desirable properties.
In particular, it may transfer properties that codes with constant
rate and sub-constant relative distance are known to have, to codes
with constant rate and constant relative distance, and even to codes
approaching the Singleton bound. We believe that this is a good ``take-home
message'' from this work.

\paragraph*{Correctable \emph{and} testable codes.}

Using the above method, it is also possible to construct improved
codes that are simultaneously locally correctable and locally testable.
This can be done by applying the distance-amplification technique
to the lifted Reed-Solomon codes of~\cite{GKS13}. The codes of~\cite{GKS13}
are both locally correctable and testable, and achieve rates that
are arbitrarily close to~$1$. Using these codes of~\cite{GKS13}
in the sub-constant relative distance regime, and combining with our
framework, we get codes of constant rate and constant relative distance
(which over large alphabets approach the Singleton bound) that are
both locally correctable and locally testable with $n^{O(1/\log\log n)}$
queries.

\paragraph*{Locally decodable codes.}

An important variant of LCCs are \textsf{locally decodable codes}
(LDCs). Those codes are defined similarly to LCCs, with the following
difference: Recall that in the definition of LCCs, the decoder gets
access to a string~$z$ which is close to a codeword~$c$, and is
required to decode a coordinate of~$c$. In the definition of~LDCs,
we view the codeword~$c$ as the encoding of some message~$x$,
and the decoder is required to decode a coordinate of~$x$. LDCs
were studied extensively in the literature, perhaps more so than LCCs
(see~\cite{Y12} for a survey). One notable fact about LDCs is that
there are constructions of LDCs with a constant query complexity and
sub-exponential length~\cite{Y08,R07,KY09,E12}.

If we restrict ourselves to \emph{linear} codes, then LDCs are a weaker
object than LCCs, since every linear LCC can be converted into an
LDC by choosing a systematic encoding map%
\footnote{This conversion will lead to an LDC with the same query complexity,
but the running time of the local decoder will be small only if the systematic encoding
map can be computed efficiently.%
}. Since the LCCs we construct in this paper are linear, all our results
apply to LDCs as well.

\paragraph*{Organization of this paper.}

We review the required preliminaries in Section~\ref{sec:Preliminaries},
construct our LCCs in Section~\ref{sec:LCCs}, and construct our
LTCs in Section~\ref{sec:LTCs}. We conclude with some open questions
in Section~\ref{sec:conclusions}.

\paragraph*{Version.}

A preliminary version of this paper appeared as~\cite{Meir-ECCC},
where the distance-amplification technique was used to construct codes
approaching the Singleton bound with query complexity $O(n^{\beta})$
(for arbitrary $\beta>0$).

\section{\label{sec:Preliminaries}Preliminaries}

All logarithms in this paper are in base~$2$. For any $n\in\N$
we denote $\left[n\right]\eqdef\left\{ 1\ldots,n\right\} $. We denote
by $\F_{2}$ the finite field of two elements. For any finite alphabet~$\Sigma$
and any pair of strings $x,y\in\Sigma^{n}$, the \textsf{relative Hamming
distance} (or, simply, \textsf{relative distance}) between $x$ and
$y$ is the fraction of coordinates on which $x$ and $y$ differ,
and is denoted by $\dist(x,y)\eqdef\left|\left\{ i\in\left[n\right]:x_{i}\ne y_{i}\right\} \right|/n$.
We have the following useful approximation.
\begin{fact}
\label{power-approximation}For every $x,y\in\R$ such that $0\le x\cdot y\le1$,
it holds that
\[
\left(1-x\right)^{y}\le1-\frac{1}{4}\cdot x\cdot y.
\]
\end{fact}
\begin{myproof}
It holds that
\[
\left(1-x\right)^{y}\le e^{-x\cdot y}\le1-\frac{1}{4}\cdot x\cdot y.
\]
The second inequality relies on the fact that $1-\frac{1}{4}\cdot x\ge e^{-x}$
for every $x\in\left(0,1\right)$, which can be proved by noting that
$1-\frac{1}{4}\cdot x=e^{-x}$ at $x=0$, and that the derivative
of $e^{-x}$ is smaller than that of $1-\frac{1}{4}\cdot x$ for every
$x\in\left(0,1\right)$. The first inequality relies on the fact that
$1-x\le e^{-x}$ for every $x\in\R$, which can be proved using similar
considerations.
\end{myproof}

\subsection{Error correcting codes}

Let $\Sigma$ be an alphabet and let $n$ be a positive integer (the
\textsf{block length}). A code is simply a subset $C\subseteq\Sigma^{n}$.
If $\F$ is a finite field and $\Sigma$ is a vector space over $\F$,
we say that a code $C\subseteq\Sigma^{n}$ is \emph{$\F$-linear} if it
is an $\F$-linear subspace of the $\F$-vector space $\Sigma^{n}$.
The \textsf{rate} of a code is the ratio $\frac{\log|C|}{\log(|\Sigma|^{n})}$,
which for $\F$-linear codes equals $\frac{\dim_{\F}(C)}{n\cdot\dim_{\F}(\Sigma)}$.

The elements of a code $C$ are called \textsf{codewords}. We say
that $C$ has \textsf{relative distance} at least $\delta$ if for every pair of
distinct codewords $c_{1},c_{2}\in C$ it holds that $\dist(c_{1},c_{2})\ge\delta$.
We will use the notation $\dist(w,C)$ to denote the relative distance
of a string $w\in\Sigma^{n}$ from~$C$, and say that $w$ is $\varepsilon$-close
(respectively, $\varepsilon$-far) to~$C$ if $\dist(w,C)<\varepsilon$
(respectively, if $\dist(w,C)\ge\varepsilon$).

An \textsf{encoding map} for $C$ is a bijection $E_{C}:\Sigma^{k}\to C$,
where $\left|\Sigma\right|^{k}=|C|$.  We say that an infinite family of codes $\left\{ C_{n}\right\} _{n}$
is \textsf{explicit} if there is a polynomial time algorithm that
computes the encoding maps of all the codes in the family. For a code $C$ of relative
distance $\delta$, a given parameter $\tau<\delta/2$, and a string
$z\in\Sigma^{n}$, the problem of decoding from $\tau$~fraction
of errors is the task of finding the unique $c\in C$ (if any) which
satisfies $\dist(c,z)\leq\tau$.

\paragraph*{Reed-Solomon codes.}

We use the following fact, which states the existence of Reed-Solomon
codes and their relevant properties.
\begin{fact}
[\label{Reed-Solomon}Reed-Solomon Codes \cite{RS60}]For every
$k,n\in\N$ such that $n\ge k$, and for every finite field $\F$
such that $\left|\F\right|\ge n$, there exists an $\F$-linear code
$\RS_{k,n}\subseteq\F^{n}$ with rate $r=k/n$, and relative distance
at least~$1-\frac{k-1}{n}>1-r$. Furthermore, $\RS_{k,n}$ has an
encoding map $E:\F^{k}\to\RS_{k,n}$ which can be computed in time
$\poly(n,\log\left|\F\right|)$, and can be decoded from up to $(1-\frac{k-1}{n})/2$
fraction of errors in time $\poly(n,\log\left|\F\right|)$. 
\end{fact}

\subsection{Locally-correctable codes}

Intuitively, a code is said to be locally correctable~\cite{BFLS91,STV01,KT00}
if, given a codeword $c\in C$ that has been corrupted by some errors,
it is possible to decode any coordinate of $c$ by reading only a
small part of the corrupted version of $c$. Formally, it is defined
as follows.
\begin{definition}
We say that a code $C\subseteq\Sigma^{n}$ is \textsf{locally correctable
from $\tau$-fraction of errors with query complexity $q$} if there
exists a randomized algorithm $A$ that satisfies the following requirements: 
\begin{itemize}
\item \textbf{Input:} $A$ takes as input a coordinate $i\in\left[n\right]$
and also gets oracle access to a string $z\in\Sigma^{n}$ that is
$\tau$-close to a codeword $c\in C$. 
\item \textbf{Output:} $A$ outputs $c_{i}$ with probability at least $\frac{2}{3}$. 
\item \textbf{Query complexity:} $A$ makes at most $q$ queries to the
oracle $z$. 
\end{itemize}
We say that the algorithm~$A$ is a \textsf{local corrector} of~$C$.
Given an infinite family of LCCs $\left\{ C_{n}\right\} _{n}$, a
\textsf{uniform local corrector }for the family is a randomized oracle
algorithm that given $n$, computes the local corrector of~$C_{n}$.
We will often be also interested in the running time of the uniform local
corrector. 
\end{definition}
\begin{remark}
The above success probability of~$\frac{2}{3}$ can be amplified
using sequential repetition, at the cost of increasing the query complexity.
Specifically, amplifying the success probability to $1-e^{-t}$ requires
increasing the query complexity by a factor of $O(t)$.
\end{remark}

\subsection{\label{sub:Locally-testable-codes}Locally-testable codes}

Intuitively, a code is said to be locally testable~\cite{FS95,RS96,GS00}
if, given a string $z\in\Sigma^{n}$, it is possible to determine
whether $z$ is a codeword of $C$, or rather far from $C$, by reading
only a small part of $z$. There are two variants of LTCs in the literature,
``weak'' LTCs and ``strong'' LTCs. From now on, we will work exclusively
with strong LTCs, since it is a simpler notion and allows us to state
a stronger result.
\begin{definition}
\label{def:LTC} We say that a code $C\subseteq\Sigma^{n}$ is \textsf{(strongly)
locally testable with query complexity} $q$ if there exists a randomized
algorithm $A$ that satisfies the following requirements: 
\begin{itemize}
\item \textbf{Input:} $A$ gets oracle access to a string $z\in\Sigma^{n}$. 
\item \textbf{Completeness:} If $z$ is a codeword of $C$, then $A$ accepts
with probability~$1$. 
\item \textbf{Soundness:} If $z$ is not a codeword of $C$, then $A$ rejects
with probability at least $\dist(z,C)$. 
\item \textbf{Query complexity:} $A$ makes at most $q$ \emph{non-adaptive}
queries to the oracle $z$. 
\end{itemize}
We say that the algorithm~$A$ is a \textsf{local tester} of~$C$.
Given an infinite family of LTCs $\left\{ C_{n}\right\} _{n}$, a
\textsf{uniform local tester }for the family is a randomized oracle
algorithm that given $n$, computes the local tester of~$C_{n}$.
Again, we will often also be interested in the running time of the uniform
local tester. 
\end{definition}

\paragraph*{A remark on amplifying the rejection probability.}

It is common to define strong LTCs with an additional parameter~$\rho$,
and have the following soundness requirement:
\begin{itemize}
\item If $z$ is not a codeword of $C$, then $A$ rejects with probability
at least $\rho\cdot$$\dist(z,C)$. 
\end{itemize}
Our definition corresponds to the special case where $\rho=1$. However,
given an LTC with $\rho<1$, it is possible to amplify $\rho$ up
to~$1$ at the cost of increasing the query complexity. Hence, we
chose to fix $\rho$ to~$1$ in our definition, which somewhat simplifies
the presentation.

The amplification of $\rho$ is performed as follows: The amplified
tester invokes the original tester~$A$ for $\frac{4}{\rho}$ times,
and accepts only if all invocations of~$A$ accept. Clearly, this
increases the query complexity by a factor of $\frac{4}{\rho}$ and
preserves the completeness property. To analyze the rejection probability,
let $z$ be a string that is not a codeword of $C$, and observe that
the amplified tester rejects it with probability at least
\begin{eqnarray*}
 &  & 1-\left(1-\rho\cdot\dist(z,C)\right)^{\frac{4}{\rho}}\\
& \ge & 1-\left(1-\frac{1}{4}\cdot\frac{4}{\rho}\cdot\rho\cdot\dist(z,C)\right) \quad \quad  \mbox{(Fact \ref{power-approximation})} \\
 & = & \dist(z,C),
\end{eqnarray*}
as required.

\subsection{\label{sub:Expander-graphs}Expander graphs}

Expander graphs are graphs with certain pseudorandom connectivity
properties. Below, we state the construction and properties that we
need. The reader is referred to~\cite{HLW06} for a survey. For a
graph $G$, a vertex $s$ and a set of vertices $T$, let $E(s,T)$
denote the set of edges that go from~$s$ into~$T$.
\begin{definition}
\label{def:sampler}Let $G=\left(U\cup V,E\right)$ be a bipartite
$d$-regular graph with $\left|U\right|=\left|V\right|=n$. We say
that $G$ is an \textsf{$(\alpha,\gamma)$-sampler} if the following
holds for every $T\subseteq V$: For at least $1-\alpha$~fraction
of the vertices $s\in U$ it holds that 
\[
\frac{\left|E(s,T)\right|}{d}-\frac{\left|T\right|}{n}\leq\gamma.
\]
\end{definition}
\begin{lemma}
\label{lem:expander-samplers}For every $\alpha,\gamma>0$ and every sufficiently large $n\in\N$
there exists a bipartite $d$-regular graph $G_{n,\alpha,\gamma}=\left(U\cup V,E\right)$
with $\left|U\right|=\left|V\right|=n$ and $d=\poly\left(\frac{1}{\alpha\cdot\gamma}\right)$
such that $G_{n,\alpha,\gamma}$ is an $\left(\alpha,\gamma\right)$-sampler.
Furthermore, there exists an algorithm that takes as inputs $n$,
$\alpha$, $\gamma$, and a vertex $w$ of $G_{n,\alpha,\gamma}$,
and computes the list of the neighbors of $w$ in $G_{n,\alpha,\gamma}$
in time $\poly(\frac{\log n}{\alpha\cdot\gamma})$.\end{lemma}
\begin{myproof}
[Proof sketch.]A full proof of Lemma~\ref{lem:expander-samplers}
requires several definitions and lemmas that we have not stated, such
as second eigenvalue, edge expansion, and the expander mixing lemma.
Since this is not the focus of this paper, we only sketch the proof
without stating those notions. The interested reader is referred to
\cite{HLW06}.

Let $\alpha$, $\gamma$ and $n$ be as in the lemma. We sketch the
construction of the graph $G\eqdef G_{n,\alpha,\gamma}$. First, observe
that it suffices to construct a strongly-explicit \emph{non-bipartite}
graph $G'$ over $n$ vertices (that is, a graph $G'$ in which the neighborhood of any given vertex is 
computable in time $\poly(\log n)$) with the desired property.
The reason is that each such graph $G'$ can be converted into a bipartite
graph~$G$ with the desired property, by taking two copies of the
vertex set of $G'$ and connecting the two copies according to the
edges in $G'$. The existence of the algorithm stated in the lemma
follows from the fact that $G'$ is strongly-explicit.

We thus focus on constructing the graph $G'$. This is done in two
steps: first, we show how to construct a strongly-explicit expander
$G''$ over $n$ vertices -- this requires a bit of work, since $n$
can be an arbitrary number, and expanders are usually constructed
for special values of~$n$. In the second step, we amplify the spectral
gap of~$G''$ by powering, and set $G'$ to be the powered graph.
We then prove that $G'$ has the desired sampling property.

\paragraph*{The first step.}

The work of~\cite{GG81} gives a strongly-explicit expander with constant degree and constant edge expansion for every
$n$ that is a square, so we only need to deal with the case in which
$n$ is not a square. Suppose that $n=m^{2}-k$, where $m^{2}$ is
the minimal square larger than~$n$, and observe that $k\le2m-1$,
which is at most $\frac{1}{2}\cdot m^{2}$ for sufficiently large~$m$.
Now, we construct an expander over $m^{2}$ vertices using \cite{GG81},
and then merge $k$ pairs of vertices. In order to maintain the regularity,
we add self-loops to all the vertices that were not merged. We set
$G''$ to be the resulting graph.

It is easy to see that $G''$ is a regular graph over $n$ vertices.
Since the merge and the addition of self-loops maintain the degree and the edge expansion
of the original expander up to a constant factor, it follows that
$G''$ is an expander with constant degree and constant edge expansion. Furthermore, it is not hard to see that $G''$
is strongly-explicit.

\paragraph*{The second step.}

Since $G''$ is an expander, and in particular has constant edge expansion,
it follows from the Cheeger inequality \cite{D84_cheeger,AM85_cheeger}
that its second-largest normalized eigenvalue (in absolute value) is some constant 
smaller than~$1$. Let us denote this normalized eigenvalue
by $\lambda$. We note that the degree and the edge expansion of $G''$, as well as $\lambda$,  are
independent of $n$.

We now construct the graph $G'$ by raising $G''$ to the
power $\log_{\lambda}\left(\sqrt{\alpha}\cdot\gamma\right)$. Observe
that $G'$ is a graph over $n$ vertices with degree $d\eqdef\poly\left(\frac{1}{\alpha\cdot\gamma}\right)$
and normalized second eigenvalue~$\sqrt{\alpha}\cdot\gamma$. It
is not hard to see that $G'$ is strongly-explicit.

\paragraph*{The sampling property.}

We prove that $G'$ has the desired sampling property. Let $T$ be
a subset of vertices of $G'$. We show that for at least $\left(1-\alpha\right)$~fraction
of the vertices $s$ of $G'$ it holds that
\[
\frac{\left|E(s,T)\right|}{d}-\frac{\left|T\right|}{n}\leq\gamma.
\]
To this end, let 
\[
S\eqdef\set{s\in U\;\bigg|\;\frac{\left|E(s,T)\right|}{d}-\frac{\left|T\right|}{n}>\gamma}.
\]
Clearly, it holds that
\[
\frac{\left|E(S,T)\right|}{d\cdot\left|S\right|}-\frac{\left|T\right|}{n}>\gamma.
\]
On the other hand, the expander mixing lemma \cite{AC88_mixing_lemma}
implies that 
\[
\frac{\left|E(S,T)\right|}{d\cdot\left|S\right|}-\frac{\left|T\right|}{n}\leq\sqrt{\alpha}\cdot\gamma\cdot\sqrt{\frac{\left|T\right|}{\left|S\right|}}.
\]
By combining the above pair of inequalities, we get
\[
\gamma<\sqrt{\alpha}\cdot\gamma\cdot\sqrt{\frac{\left|T\right|}{\left|S\right|}}
\]
\[
\left|S\right|<\alpha\cdot\left|T\right|\le\alpha\cdot n,
\]
as required.
\end{myproof}

\section{\label{sec:LCCs}LCCs with sub-polynomial query complexity}

In this section, we prove the following theorem on LCCs, which immediately
implies Theorem~\ref{main-thm-LCCs} from the introduction.
\begin{theorem}
[\label{re-main-thm-LCCs}Main LCC theorem]For every $r \in (0,1)$,
there exists an explicit infinite family of $\F_2$-linear codes $\left\{ C_{n}\right\} _{n}$
satisfying: 
\begin{enumerate}
\item $C_{n}$ has block length $n$, rate at least $r$, and relative distance
at least $1-r-o(1)$.
\item $C_{n}$ is locally correctable from $\frac{1-r-o(1)}{2}$~fraction
of errors with query complexity $\exp(\sqrt{\log n\cdot\log\log n})$.
\item The alphabet of $C_{n}$ is a vector space $\Sigma_{n}$ over $\F_{2}$,
such that $\left|\Sigma_{n}\right|\leq\exp\left(\exp(\sqrt{\log n\cdot\log\log n})\right)$.
\end{enumerate}
Furthermore, the family $\left\{ C_{n}\right\} _{n}$ has a uniform
local corrector that 
runs in time $\exp(\sqrt{\log n\cdot\log\log n})$.
\end{theorem}
We note that the existence of binary LCCs (Theorem~\ref{thm:binary-LCCs})
also follows from Theorem~\ref{re-main-thm-LCCs}: In order to construct
the binary LCCs, we concatenate the codes of Theorem~\ref{re-main-thm-LCCs}
with any asymptotically good inner binary code that has efficient
encoding and decoding algorithms. The local corrector of the binary
LCCs will emulate the original local corrector, and whenever the latter
queries a symbol, the binary local corrector will emulate this query
by decoding the corresponding codeword of the inner code. Since such
constructions are standard (see~\cite{KSY14}), we do not provide the
full details.

The proof of Theorem \ref{re-main-thm-LCCs} has two steps. In the
first step, we give a transformation that amplifies the fraction of
errors from which an LCC can be corrected -- this step follows the
distance amplification of~\cite{AL96_codes}. In the second step,
we construct a locally-correctable code $W_{n}$ with the the desired
query complexity but that can only be corrected from a sub-constant
fraction of errors. Finally, we construct the code $C_{n}$ by applying
the distance amplification to $W_{n}$. Those two steps are formalized
in the following pair of lemmas, which are proved in Sections \ref{subsec:LCC-transform}
and~\ref{subsec:LCC-sub-poly} respectively.
\begin{lemma}
\label{lem:LCC-transform}Suppose that there exists a code $W$ that is locally correctable
from $\tW$~fraction of errors with query complexity~$q$, such
that:
\begin{itemize}
\item $W$ has rate $\rW$.
\item $W$ is $\F_{2}$-linear 
\end{itemize}
Then, for every $0<\tau<\frac{1}{2}$ and $0< \varepsilon < 1$, there exists a code $C$ that
is locally correctable from $\tau$~fraction of errors with query
complexity~$q\cdot\poly(1/(\varepsilon\cdot\tW))$, such that:
\begin{itemize}
\item $\left|C\right|=\left|W\right|$.
\item $C$ has relative distance at least~$2\cdot\tau$, and rate at least
$\rW\cdot(1-2\cdot\tau-\varepsilon)$.
\item Let $\Lambda$ denote the alphabet of~$W$. Then, the alphabet of~$C$
is $\Sigma\eqdef\Lambda^{p}$ for some $p=\poly(1/(\varepsilon\cdot\nolinebreak\tW))$.
\item $C$ is $\F_{2}$-linear.
\end{itemize}
Furthermore,
\begin{itemize}
\item There is a polynomial time algorithm that computes a bijection from
every code~$W$ to the corresponding code~$C$, given $\rW$, $\tW$,
$r$,~$\varepsilon$ and~$\Lambda$.
\item There is an oracle algorithm that when given black box access to the
local corrector of any code~$W$, and given also $\rW$, $\tW$,
$r$, $\varepsilon$, $\Lambda$, computes the local corrector of
the corresponding code~$C$. The resulting local corrector of~$C$
runs in time that is polynomial in the running time of the local corrector
of~$W$ and in $1/\tW$, $1/\varepsilon$ and $\log(n_W)$ where $n_W$ is the block length of $W$.
\end{itemize}
\end{lemma}

\begin{lemma}
\label{high-rate-LCC}There exists an explicit infinite family of
$\F_{2}$-linear codes $\left\{ W_{n}\right\} _{n}$ satisfying:
\begin{enumerate}
\item $W_{n}$ has block length $n$, rate at least $1-\frac{1}{\log n}$,
and relative distance at least $\Omega\left(\sqrt{\frac{\log\log n}{\log^{3}n}}\right)$.
\item $W_{n}$ is locally correctable from $\Omega\left(\sqrt{\frac{\log\log n}{\log^{3}n}}\right)$
fraction of errors with query complexity $\exp(\sqrt{\log n\cdot\log\log n})$.
\item The alphabet of $W_{n}$ is a vector space $\Lambda_n$ over $\F_2$, such that  $|\Lambda_n| \leq \exp\left(\exp(\sqrt{\log n\cdot\log\log n})\right)$.
\end{enumerate}
Furthermore, the family $\left\{ W_{n}\right\} _{n}$ has a uniform
local corrector that runs in time $\exp(\sqrt{\log n\cdot\log\log n})$.\end{lemma}
\begin{myproof}
[Proof of Theorem \ref{re-main-thm-LCCs}.]We construct the family
$\left\{ C_{n}\right\} _{n}$ by applying Lemma~\ref{lem:LCC-transform}
to the family~$\left\{ W_{n}\right\} _{n}$ of Lemma~\ref{high-rate-LCC}
with $\tW=\Omega\left(\sqrt{\frac{\log\log n}{\log^{3}n}}\right)$,
$\varepsilon=\frac{1}{\log n}$, and 
\[
\tau=\frac{1}{2}\cdot\left(1-\frac{r}{1-\frac{1}{\log n}}-\varepsilon\right)=\frac{1}{2}.\left(1-r-O\left(\frac{1}{\log n}\right)\right).
\]
It is easy to see that $C_{n}$ has the required rate,
relative distance and alphabet size, and that it can be locally corrected from the required
fraction of errors with the required query complexity. The family
$\left\{ C_{n}\right\} _{n}$ is explicit with the required running time due to the first item in
the ``furthermore'' part of Lemma~\ref{lem:LCC-transform}, and
has a uniform local corrector due to the second item of that part.\end{myproof}
\begin{remark}
In Lemma~\ref{lem:LCC-transform} above, we chose to assume that
$W$ is $\F_{2}$-linear for simplicity. More generally, if $W$ is
$\F$-linear for any finite field~$\F$, then $C$ is $\F$-linear
as well. Furthermore, the lemma also works if $W$ is not $\F$-linear
for any field~$\F$, in which case $C$ is not guaranteed to be $\F$-linear
for any~field~$\F$.
\end{remark}

\subsection{\label{subsec:LCC-transform}Proof of Lemma \ref{lem:LCC-transform}}

\subsubsection{\label{sub:LCC-overview}Overview}

Let $0<\tau<\frac{1}{2}$. Our goal is to construct a code $C$ that
can be locally corrected from a fraction of errors at most~$\tau$.
The idea of the construction is to combine the LCC $W$ with a Reed-Solomon
code to obtain a code $C$ that enjoys ``the best of both worlds'':
both the local correctability of~$W$ and the good error correction
capability of Reed-Solomon. We do it in two steps: first, we construct
a code $C'$ which can be corrected from $\tau$~fraction of \emph{random}
errors. Then, we augment $C'$ to obtain a code~$C$ that can be
corrected from $\tau$~fraction of \emph{adversarial} errors.

We first describe the construction of $C'$. To this end, we describe
a bijection from $W$ to $C'$. Let $w$ be a codeword of~$W$. To
obtain the codeword $c'\in C'$ that corresponds to~$w$, we partition
$w$ into blocks of length $b$ (to be determined later), and encode
each block with a Reed-Solomon code $\RS_{b,d}$. We choose the relative
distance of $\RS_{b,d}$ to be $2\cdot\tau+\varepsilon$, so its rate
is $1-2\cdot\tau-\varepsilon$ and the rate of $C'$ is indeed $\rW\cdot\left(1-2\cdot\tau-\varepsilon\right)$,
as required.

We now claim that if one applies to a codeword~$c'\in C'$ a noise
that corrupts each coordinate with probability~$\tau$, then the
codeword $c'$ can be recovered from its corrupted version with high
probability. To see it, first observe that with high probability,
almost all the blocks of~$c'$ have at most $\tau+\frac{\varepsilon}{2}$~fraction
of corrupted coordinates. Let us call those blocks ``good blocks'',
and observe that the good blocks can be corrected by decoding them
to the nearest codeword of~$\RS_{b,d}$ (since $\tau+\frac{\varepsilon}{2}$
is half the relative distance of $\RS_{b,d}$). Next, observe that
if $b$ is sufficiently large, the fraction of ``good blocks'' is
at least $1-\tW$, and hence we can correct the remaining $\tW$~fraction
of errors using the decoding algorithm of~$W$. It follows that $C'$
can be corrected from $\tau$~fraction of random errors, as we wanted.

Next, we show how to augment $C'$ to obtain a code~$C$ that is
correctable from adversarial errors. This requires two additional
ideas. The first idea to apply a permutation that is ``pseudorandom''
in some sense to the coordinates of~$C'$. The ``pseudorandom''
permutation is determined by the edges of an expander graph (see Section~\ref{sub:Expander-graphs}).
This step is motivated by the hope that, after the adversary decided
which coordinates to corrupt, applying the permutation to the coordinates
will make the errors behave pseudorandomly. This will allow the above
analysis for the case of random errors to go through.

Of course, on its own, this idea is doomed to fail, since the adversary
can take the permutation into account when he chooses where to place
the errors. Here the second idea comes into play: after applying the
permutation to the coordinates of~$C'$, we will increase the alphabet
size of the code, packing each block of symbols into a new big symbol.
The motivation for this step is that increasing the alphabet size
restricts the freedom of the adversary in choosing the pattern of
errors. Indeed, we will show that after the alphabet size is increased,
applying the permutation to the coordinates of the code makes the
errors behave pseudorandomly. This allows us to prove that the code
can be decoded from $\tau$~fraction of errors, as we wanted.

\subsubsection{\label{sub:LCC-construction}The construction of~$C$}

\paragraph*{Choosing the parameters.}

Let $W$, $\rW$, $\tW$, $r$, $\varepsilon$, and $\Lambda$ be
as in Lemma~\ref{lem:LCC-transform}. Let $\left\{ G_{n}\right\} _{n}$
be an infinite family of $(\tW,\frac{1}{2}\cdot\varepsilon)$-samplers
as in Theorem~\ref{lem:expander-samplers}, and let $d$ be their
degree.

Recall that we assumed that $W$ is $\F_{2}$-linear, so $\left|\Lambda\right|$
is a power of~$2$. Let $\F$ be an extension field of~$\F_{2}$,
whose size is the minimal power of $\left|\Lambda\right|$ that is
at least $d$. Let $\RS_{b,d}$ be a Reed-Solomon code over $\F$
with relative distance $2\cdot\tau+\varepsilon$, rate $1-2\cdot\tau-\varepsilon$,
and block length~$d$. 

Let $\nW$ be the block length of~$W$, and let $t$ be such that $\left|\F\right|=\left|\Lambda\right|^{t}$. The block length of~$C$ will be $n\eqdef\frac{n_{W}}{b\cdot t}$,
and its alphabet will be $\Sigma\eqdef\F^{d}$. Here, we assume that $\nW$ is divisible by
$b\cdot t$. If $\nW$ is not divisible by~$b\cdot t$, we consider two cases:
\begin{itemize}
\item if $\nW > b\cdot t / \varepsilon$, we increase $\nW$ to the next multiple of $b\cdot t$ by padding the
codewords of $W$ with additional zero coordinates. This decreases
the rate of $W$ by at most $\varepsilon$, which essentially does not affect our results.

\item Otherwise, we set $C$ to be any Reed-Solomon code with blocklength $\nW$, relative distance $2 \cdot \tau$, and rate $1 - 2 \cdot \tau$. Observe that such a Reed-Solomon is locally correctable from $\tau$~fraction of errors with query complexity
$$\nW \le b\cdot t / \varepsilon = \poly(1/ (\varepsilon \cdot \tW)),$$
which satisfies our requirements.
\end{itemize}

\paragraph*{A bijection from $W$ to~$C$.}

We construct the code $C$ by describing a bijection from $W$ to~$C$.
Given a codeword $w\in W$, one obtains the corresponding codeword
$c\in C$ as follows: 
\begin{itemize}
\item Partition $w$ into $n\eqdef\frac{\nW}{b\cdot t}$ blocks of length
$b\cdot t$. We view each of those blocks as a vector in~$\F^{b}$,
and encode it via the code~$\RS_{b,d}$. Let us denote the resulting
string by $c'\in\F^{n\cdot d}$ and the resulting codewords of $\RS_{b,d}$
by $B_{1},\ldots,B_{n}\in\F^{d}$. 
\item Next, we apply a ``pseudorandom'' permutation to the coordinates
of~$c'$ as follows: Let $G_{n}$ be the graph from the infinite
family above and let $U=\left\{ u_{1},\ldots,u_{n}\right\} $ and
$V=\left\{ v_{1},\ldots,v_{n}\right\} $ be the left and right vertices
of~$G_{n}$ respectively. For each $i\in\left[n\right]$ and $j\in\left[d\right]$,
we write the $j$-th symbol of $B_{i}$ on the $j$-th edge of $u_{i}$.
Then, we construct new blocks $S_{1},\ldots,S_{n}\in\F^{d}$,
by setting the $j$-th symbol of $S_{i}$ to be the symbol written
on the $j$-th edge of $v_{i}$. 
\item Finally, we define the codeword $c$ of~$C\subseteq\Sigma^{n}$ as
follows: the $i$-th coordinate $c_{i}$ is the block $S_{i}$, reinterpreted
as a symbol of the alphabet $\Sigma\eqdef\F^{d}$. We choose $c$
to be the codeword in $C$ that corresponds to the codeword $w$ in
$W$. 
\end{itemize}
This concludes the definition of the bijection. It is not hard to
see that this bijection can be computed in polynomial time, and that
the code~$C$ is $\F_{2}$-linear. Furthermore, $\Sigma = \F^d = \Lambda^{t \cdot d}$ where $d \cdot t \leq d \log d = \poly(1/(\varepsilon \cdot \tau_W))$. The rate of $C$ is
\begin{eqnarray*}
\frac{\log|C|}{n\cdot\log|\Sigma|} & = & \frac{\log|W|}{n\cdot d\cdot\log\left|\F\right|}\\
 & = & \frac{\rW\cdot\log\left|\Lambda^{\nW}\right|}{n\cdot d\cdot\log\left|\F\right|}\\
 & = & \rW\cdot\frac{\nW}{n}\cdot\frac{1}{d}\cdot\frac{\log|\Lambda|}{\log|\F|}\\
 & = & \rW\cdot\left(b\cdot t\right)\cdot\frac{1-2\cdot\tau-\varepsilon}{b}\cdot\frac{1}{t}\\
 & = & \rW\cdot(1-2\cdot\tau-\varepsilon),
\end{eqnarray*}
as required. The relative distance of~$C$ is at least $2\cdot\tau$
-- although this could be proved directly, it also follows immediately
from the fact that~$C$ is locally correctable from $\tau$ fraction
of errors, which is proved in the next section.

\subsubsection{\label{sub:LCC-local-correctability}Local correctability}

In this section, we complete the proof of Lemma \ref{lem:LCC-transform}
by proving that $C$ is locally correctable from $\tau$~fraction
of errors with query complexity $\poly(d)\cdot q$. To this end,
we describe a local corrector~$A$. The algorithm~$A$ is based
on the following algorithm $A_{0}$, which locally corrects coordinates
of $W$ from a corrupted codeword of~$C$. 
\begin{lemma}
\label{intermediate-decoding}There exists an algorithm $A_{0}$ that
satisfies the following requirements: 
\begin{itemize}
\item \textbf{Input:} $A_{0}$ takes as input a coordinate $i\in\left[\nW\right]$,
and also gets oracle access to a string $z\in\Sigma^{n}$ that is
$\tau$-close to a codeword $c\in C$.
\item \textbf{Output:} Let $w^{c}$ be the codeword of $W$ from which $c$
was generated. Then, $A_{0}$ outputs $w_{i}^{c}$ with probability
at least $1-\frac{1}{3\cdot b\cdot t\cdot d}$. 
\item \textbf{Query complexity:} $A_{0}$ makes $\poly(d)\cdot q$ queries
to the oracle $z$.
\end{itemize}
\end{lemma}
Before proving Lemma~\ref{intermediate-decoding}, we show how to
construct the algorithm~$A$ given the algorithm $A_{0}$. Suppose
that the algorithm~$A$ is given oracle access to a string~$z$
that is $\tau$-close to a codeword $c\in C$, and a coordinate~$i\in\left[n\right]$.
The algorithm is required to decode~$c_{i}$. Let $w^{c}\in\Lambda^{\nW}$
be the codeword of $W$ from which $c$ was generated, and let $B_{1}^{c},\ldots,B_{n}^{c}$
and $S_{1}^{c},\ldots,S_{n}^{c}$ be the corresponding blocks.

In order to decode~$c_{i}$, the algorithm~$A$ should decode each
of the symbols in the block~$S_{i}^{c}\in\F^{d}$. Let $u_{j_{1}},\ldots,u_{j_{d}}$
be the neighbors of $v_{i}$ in the graph~$G_{n}$. Each symbol of
the block $S_{i}^{c}$ belongs to one of the blocks $B_{j_{1}}^{c},\ldots,B_{j_{d}}^{c}$,
and therefore it suffices to retrieve the latter blocks. Now, each
block~$B_{j_{h}}^{c}$ is the encoding via $\RS_{b,d}$ of $b\cdot t$
symbols of $w^{c}$ (in the alphabet~$\Lambda$). The algorithm $A$
invokes the algorithm~$A_{0}$ to decode each of those $b\cdot t$
symbols of~$w^{c}$, for each of the blocks $B_{j_{1}}^{c},\ldots,B_{j_{d}}^{c}$.
By the union bound, the algorithm $A_{0}$ decodes all those $b\cdot t\cdot d$
symbols of $w^{c}$ correctly with probability at least $1-b\cdot t\cdot d\cdot\frac{1}{3\cdot b\cdot t\cdot d}=\frac{2}{3}$.
Whenever that happens, the algorithm~$A$ retrieves the blocks $B_{j_{1}}^{c},\ldots,B_{j_{d}}^{c}$
correctly, and therefore computes the block~$S_{i}^{c}$ correctly.
This concludes the construction of the algorithm~$A$. Note that
the query complexity of~$A$ is larger than that of~$A_{0}$ by
a factor of at most~$b\cdot t\cdot d$, and hence it is at most~$\poly(d)\cdot q$.
It remains to prove Lemma~\ref{intermediate-decoding}. 
\begin{myproof}
[Proof of Lemma~\ref{intermediate-decoding}.]Let $\AW$ be the
local corrector of the code~$W$. By amplification, we may assume
that $\AW$ errs with probability at most $\frac{1}{3\cdot b\cdot t\cdot d}$,
and this incurs a factor of at most $\poly(d)$ to its query complexity.

Suppose that the algorithm~$A_{0}$ is invoked on a string~$z\in\Sigma^{n}$
and a coordinate~$i\in\left[\nW\right]$. The algorithm $A_{0}$
invokes the algorithm~$\AW$ to retrieve the coordinate~$i$, and
emulates $\AW$ in the natural way: Recall that $\AW$ expects to
be given access to a corrupted codeword of~$W$, and makes queries
to it. Whenever $\AW$ makes a query to a coordinate $\iW\in\left[\nW\right]$,
the algorithm $A_{0}$ performs the following steps. 
\begin{enumerate}
\item $A_{0}$ finds the block $B_{l}$ to which the coordinate $\iW$ belongs.
Formally, $l\eqdef\left\lceil \iW/(b\cdot t)\right\rceil $. 
\item $A_{0}$ finds the neighbors of the vertex $u_{l}$ in~$G_{n}$.
Let us denote those vertices by $v_{j_{1}},\ldots,v_{j_{d}}$. 
\item $A_{0}$ queries the coordinates $j_{1},\ldots j_{d}$, thus obtaining
the blocks $S_{j_{1}},\ldots,S_{j_{d}}$. 
\item $A_{0}$ reconstructs the block $B_{l}$ by reversing the permutation
of $G_{n}$ on $S_{j_{1}},\ldots,S_{j_{d}}$. 
\item $A_{0}$ attempts to decode $B_{l}$ by applying an efficient decoding
algorithm of Reed-Solomon.
\item Suppose that the decoding succeeded and returned a codeword of $\RS_{b,d}$
that is $\left(\tau+\frac{\varepsilon}{2}\right)$-close to $B_{l}$.
Then, $A_{0}$ retrieves the value of the $\iW$-th coordinate of~$w^{c}$
from the latter codeword, and feeds it to $\AW$ as an answer to its
query. 
\item Otherwise, $A_{0}$ feeds $0$ as an answer to the query of~$\AW$.
\end{enumerate}
When the algorithm $\AW$ finishes running, the algorithm~$A_{0}$
finishes and returns the output of~$\AW$. It is not hard to see
that the query complexity of~$A_{0}$ is at most $d$ times the query
complexity of~$\AW$, and hence it is at most $\poly(d)\cdot q$. It remains
to show that $A_{0}$ succeeds in decoding from $\tau$~fraction
of errors with probability at least $1-\frac{1}{3\cdot b\cdot t\cdot d}$.

Let $z\in\Sigma^{n}$ be a string that is $\tau$-close to a codeword
$c\in C$. Let $w^{c}\in\Lambda^{\nW}$ be the codeword of $W$ from
which $c$ was generated, and let $B_{1}^{c},\ldots,B_{n}^{c}$ and
$S_{1}^{c},\ldots,S_{n}^{c}$ be the corresponding blocks. We also
use the following definitions: 
\begin{enumerate}
\item Let $S_{1}^{z},\ldots,S_{n}^{z}\in\F^{d}$ be the blocks that correspond
to the symbols of $z$. 
\item Let $B_{1}^{z},\ldots,B_{n}^{z}$ be the blocks that are obtained
from $S_{1}^{z},\ldots,S_{n}^{z}$ by reversing the permutation. 
\item Define blocks ${B_{1}^{z}}',\ldots,{B_{n}^{z}}'$ as follows: if $B_{i}^{z}$
is $\left(\tau+\frac{\varepsilon}{2}\right)$-close to $\RS_{b,d}$,
then ${B_{i}^{z}}'$ is the nearest codeword of $\RS_{b,d}$. Otherwise,
${B_{i}^{z}}'$ is the all-zeroes block.
\item Let $w^{z}\in\Lambda^{\nW}$ be the string that is obtained by extracting
the coordinates of $w$ from each of the codewords ${B_{1}^{z}}',\ldots,{B_{n}^{z}}'$.
\end{enumerate}
It is easy to see that $A_{0}$ emulates the action of $\AW$ on $w^{z}$.
Therefore, if we prove that $w^{z}$ is $\tW$-close to $w^{c}$,
we will be done. In order to do so, it suffices to prove that for
at least $1-\tW$ fraction of the blocks $B_{l}^{z}$, it holds that
$B_{l}^{z}$ is $\left(\tau+\frac{\varepsilon}{2}\right)$-close to
$B_{l}^{c}$. 

To this end, let $J$ be the set of coordinates on which $z$ and
$c$ differ. In other words, for every $j\in J$ it holds that $S_{j}^{z}\ne S_{j}^{c}$.
By assumption, $|J|\le\tau\cdot n$. Now, observe that since $G_{n}$
is a $\left(\tW,\frac{1}{2}\cdot\varepsilon\right)$-sampler, it holds
that for at least $\left(1-\tW\right)$~fraction of the vertices
$u_{l}$ of~$G_n$, there are at most $\left(\tau+\frac{\varepsilon}{2}\right)\cdot d$
edges between $u_{l}$ and~$J$. For each such $u_{l}$, it holds
that $B_{u_{l}}^{z}$ is $\left(\tau+\frac{\varepsilon}{2}\right)$-close
to $B_{u_{l}}^{c}$, and this concludes the proof.
\end{myproof}
It can be verified that the local correctors $A_{0}$ and~$A$ can
be implemented efficiently with black box access to~$\AW$, as required
by the second item in the ``furthermore'' part of the lemma.

\subsection{\label{subsec:LCC-sub-poly}Proof of Lemma \ref{high-rate-LCC}}

In this section we prove Lemma \ref{high-rate-LCC}, restated below.
\begin{replemma}{\ref{high-rate-LCC}}
There exists an explicit infinite family of
$\F_{2}$-linear codes $\left\{ W_{n}\right\} _{n}$ satisfying:
\begin{enumerate}
\item $W_{n}$ has block length $n$, rate at least $1-\frac{1}{\log n}$,
and relative distance at least $\Omega\left(\sqrt{\frac{\log\log n}{\log^{3}n}}\right)$.
\item $W_{n}$ is locally correctable from $\Omega\left(\sqrt{\frac{\log\log n}{\log^{3}n}}\right)$
fraction of errors with query complexity $\exp(\sqrt{\log n\cdot\log\log n})$.
\item The alphabet of $W_{n}$ is a vector space $\Lambda_n$ over $\F_2$, such that  $|\Lambda_n| \leq \exp\left(\exp(\sqrt{\log n\cdot\log\log n})\right)$.
\end{enumerate}
Furthermore, the family $\left\{ W_{n}\right\} _{n}$ has a uniform
local corrector that runs in time $\exp(\sqrt{\log n\cdot\log\log n})$.
\end{replemma}

For the proof of Lemma \ref{high-rate-LCC} we use the multiplicity codes of \cite{KSY14}, in a specialized sub-constant
relative distance regime.

\begin{lemma}
[{\label{lem:multparam}\cite[Lemma 3.5]{KSY14}}] Let $\F$ be any finite field. Let $s,d,m$ be positive integers. Let $M$ be the
multiplicity code of order $s$ evaluations of degree $d$ polynomials
in $m$ variables over $\F$. Then $M$ has block length
$\left|\F\right|^{m}$, relative distance at least $\delta\eqdef1-\frac{d}{s\cdot\left|\F\right|}$
and rate $\frac{{d+m \choose m}}{{s+m-1 \choose m}\cdot\left|\F\right|^{m}}$,
which is at least 
\[
\left(\frac{s}{m+s}\right)^{m}\cdot\left(\frac{d}{s\cdot\left|\F\right|}\right)^{m}\geq\left(1-\frac{m^{2}}{s}\right)\cdot(1-\delta)^{m}.
\]
The alphabet of~$C$ is $\F^{\binom{m+s-1}{m}}$, and $C$ is $\F$-linear.
Furthermore, there is $\poly\left(\F^m,\binom{m+s-1}{m}\right)$ time algorithm that computes an
encoding map of~$M$ given $s$, $d$, $m$, and $\F$.
\end{lemma}

\begin{lemma}
[{\label{lem:multlccparam}\cite[Lemma 3.6]{KSY14}}]Let $M$ be
the multiplicity code as above. Let $\delta=1-\frac{d}{s\cdot\left|\F\right|}$
be a lower bound for the relative distance of $M$. Suppose $\left|\F\right|\geq\max\{10\cdot m,\frac{d+6\cdot s}{s},12\cdot(s+1)\}$.
Then $M$ is locally correctable from $\delta/10$ fraction of errors
with query complexity $O(s^{m}\cdot\left|\F\right|)$.
\end{lemma}
As discussed in Section 4.3 of~\cite{KSY14},
this local corrector can be implemented to have running time $\poly(\left|\F\right|,s^{m})$
over fields of constant characteristic. In fact,~\cite{multcodesurvey}
shows that the query complexity and running time for local correcting
multiplicity codes can be further reduced to $\left|\F\right|\cdot O\left((\frac{1}{\delta})^{m}\right)$
queries, but this does not lead to any noticeable improvement for
our setting. 

We now prove Lemma \ref{high-rate-LCC}.

\begin{myproof}
Let $n\in\N$ be a codeword
length. We set the code $W_{n}$ to be a multiplicity code with the
following parameters. We choose $\F$ to be a field of size $2^{\sqrt{\log n\cdot\log\log n}}$,
and choose $m=\sqrt{\frac{\log n}{\log\log n}}$. Note that indeed
$\left|\F\right|^{m}=n$. We choose $s=2\cdot m^{2}\cdot\log n$.
Let $\delta=\frac{1}{2\cdot m\cdot\log n}$ (this will be a lower
bound on the relative distance of the code) and choose the degree
of the polynomials to be $d=s\cdot\left|\F\right|\cdot(1-\delta)$.

It can be verified that the relative distance of the code is at least
$\delta\ge\Omega\left(\sqrt{\frac{\log\log n}{\log^{3}n}}\right)$.
The rate of the code is at least
\[
\left(1-\frac{m^{2}}{s}\right)\cdot(1-\delta)^{m}\geq\left(1-\frac{1}{2\cdot\log n}\right)\left(1-\frac{1}{2\cdot m\cdot\log n}\right)^{m}\geq1-\frac{1}{\log n},
\]
as required. The alphabet size is 
\begin{eqnarray*}
\left|\F\right|^{{m+s-1 \choose m}} & \le & \exp\left(\sqrt{\log n\cdot\log\log n}\cdot s^{m}\right)\\
 & = & \exp\left(\sqrt{\log n\cdot\log\log n}\cdot\left(\frac{\log^{2}n}{\log\log n}\right)^{\sqrt{\frac{\log n}{\log\log n}}}\right)\\
 & = & \exp\left(\exp\left(\sqrt{\log n\cdot\log\log n}\right)\right).
\end{eqnarray*}
Moreover, the alphabet is a vector space over $\F$ and hence in particular
over $\F_{2}$ (since we chose the size of $\F$ to be a power of~$2$).
The code $W_{n}$ is $\F$-linear and in particular $\F_{2}$-linear.

By Lemma~\ref{lem:multlccparam}, $W_{n}$ is locally correctable
from $\frac{1}{10}\cdot\delta\ge\Omega\left(\sqrt{\frac{\log\log n}{\log^{3}n}}\right)$
fraction of errors with query complexity
\[
O(s^{m}\cdot\left|\F\right|)\le O\left(\frac{\log^{2}n}{\log\log n}\right)^{\sqrt{\frac{\log n}{\log\log n}}}\cdot2^{\sqrt{\log n\cdot\log\log n}}=2^{O(\sqrt{\log n\cdot\log\log n})},
\]
as required. Finally, the fact that the family $\left\{ W_{n}\right\} _{n}$
is explicit follows from the ``furthermore'' part of Lemma~\ref{lem:multparam},
and the fact that it has an efficient uniform local corrector with the required running time follows
from the discussion after Lemma~\ref{lem:multlccparam}.
\end{myproof}

\subsection{LDCs}

As remarked earlier, by choosing a systematic encoding map, linear
LCCs automatically give LDCs with the same rate, relative distance, and
query complexity. The running time of the local decoding algorithm
will be essentially the same as the running time of the local correction
algorithm, provided that the systematic encoding map can be computed
efficiently. Using the fact that multiplicity codes have an efficiently
computable systematic encoding map~\cite{Kop-mult-list}, it is easy
to check that the codes we construct above also have an efficiently
computable systematic encoding map. Thus we get LDCs with the same
parameters as our LCCs. 

\section{\label{sec:LTCs}LTCs with sub-polynomial query complexity}

In this section, we prove the following theorem on LTCs, which immediately
implies Theorem~\ref{main-thm-LTCs} from the introduction.
\begin{theorem}
[\label{re-main-thm-LTCs}Main LTC theorem]For every $r\in(0,1)$,
there exists an explicit infinite family of $\F_2$-linear codes $\left\{ C_{n}\right\} _{n}$
satisfying: 
\begin{enumerate}
\item $C_{n}$ has block length $n$, rate at least $r$, and relative distance
at least $1-r-o(1)$.
\item $C_{n}$ is locally testable with query complexity $\exp(\sqrt{\log n\cdot\log\log n})$.
\item The alphabet of $C_{n}$ is a vector space $\Sigma_{n}$ over $\F_{2}$,
such that $\left|\Sigma_{n}\right|\leq\exp(\exp\left(\sqrt{\log n\cdot\log\log n}\right))$.
\end{enumerate}
Furthermore, the family $\left\{ C_{n}\right\} _{n}$ has a uniform
local tester that 
runs in time $\exp(\sqrt{\log n\cdot\log\log n)})$.
\end{theorem}
We note that the existence of binary LTCs (Theorem~\ref{thm:binary-LTCs})
also follows from Theorem~\ref{re-main-thm-LTCs}: In order to construct
the binary LTCs, we concatenate the codes of Theorem~\ref{re-main-thm-LTCs}
with any asymptotically good inner binary code that has efficient
encoding and decoding algorithms. The local tester of the binary LTCs will emulate
the original local tester, and whenever the latter queries a symbol,
the binary local tester will emulate this query by reading the corresponding
codeword of the inner code. If this string is not a legal codeword,
the binary tester will reject, and otherwise it will decode the symbol
and feed it to the original tester. Since such constructions are standard, we do not provide the full details.

The proof of Theorem \ref{re-main-thm-LTCs} has two steps. In the
first step, we give a transformation that amplifies the relative distance
of an LTC -- this step follows the distance amplification of~\cite{AL96_codes}.
In the second step, we construct a locally-testable code $W_{n}$
with the desired query complexity but that has sub-constant relative
distance. Finally, we construct the code $C_{n}$ by applying the
distance amplification to $W_{n}$. Those two steps are formalized
in the following pair of lemmas, which are proved in Sections \ref{lem:LTC-transform}
and~\ref{high-rate-LTC} respectively.
\begin{lemma}
\label{lem:LTC-transform} Suppose that there exists a code $W$ with relative distance~$\dW$
that is locally testable with query complexity~$q$ such that:
\begin{itemize}
\item $W$ has rate $\rW$.
\item $W$ is $\F_{2}$-linear.
\end{itemize}
Then, for every $0<\delta, \varepsilon<1$, there exists a code $C$ with relative
distance at least ~$\delta$ that is locally testable with query complexity~$q\cdot\poly(1/(\varepsilon\cdot\dW))$,
such that:
\begin{itemize}
\item $\left|C\right|=\left|W\right|$.
\item $C$ has rate at least $\rW\cdot(1-\delta-\varepsilon)$.
\item Let $\Lambda$ denote the alphabet of~$W$. Then, the alphabet of~$C$
is $\Sigma\eqdef\Lambda^{p}$ for some $p=\poly(1/(\varepsilon\cdot\nolinebreak\dW))$.
\item $C$ is $\F_{2}$-linear.
\end{itemize}
Furthermore,
\begin{itemize}
\item There is a polynomial time algorithm that computes a bijection from
every code~$W$ to the corresponding code~$C$, given $\rW$, $\dW$,
$r$,~$\varepsilon$ and~$\Lambda$.
\item There is an oracle algorithm that when given black box access to the
local tester of any code~$W$, and given also $\rW$, $\dW$, $r$,
$\varepsilon$, $\Lambda$, and the block length of~$W$, computes
the local tester of the corresponding code~$C$. The resulting local
tester of~$C$ runs in time that is polynomial in the running time
of the local tester of~$W$ and in $1/\dW$, $1/\varepsilon$ and $\log(n_W)$ where $n_W$ is the block length of $W$.
\end{itemize}
\end{lemma}

\begin{lemma}
\label{high-rate-LTC}There exists an explicit infinite family of
$\F_{2}$-linear codes $\left\{ W_{n}\right\} _{n}$ satisfying: 
\begin{enumerate}
\item $W_{n}$ has block length $n$, rate at least $1-\frac{1}{\log n}$,
and relative distance at least $\exp(-\sqrt{\log n\cdot\log\log n})$.
\item $W_{n}$ is locally testable with query complexity $\exp(\sqrt{\log n\cdot\log\log n})$.
\item The alphabet of $W_{n}$ is a vector space $\Lambda_n$ over $\F_2$, such that 
$|\Lambda_n| \leq \exp\left(\sqrt{\log n\cdot\log\log n}\right)$.
\end{enumerate}
Furthermore, the family $\left\{ W_{n}\right\} _{n}$ has a uniform
local tester that runs in time $\exp(\sqrt{\log n\cdot\log\log n})$.\end{lemma}
\begin{myproof}
[Proof of Theorem \ref{re-main-thm-LTCs}.]We construct the family
$\left\{ C_{n}\right\} _{n}$ by applying Lemma~\ref{lem:LTC-transform}
to the family~$\left\{ W_{n}\right\} _{n}$ of Lemma~\ref{high-rate-LTC}
with $\dW=2^{-O(\sqrt{\log n\cdot\log\log n})}$, $\varepsilon=\frac{1}{\log n}$
and
\[
\delta=1-\frac{r}{1-\frac{1}{\log n}}-\varepsilon=1-r-O\left(\frac{1}{\log n}\right).
\]
It is easy to see that $C_{n}$ has
the required rate, relative distance and alphabet size, and that it can be locally
tested with the required query complexity. The family $\left\{ C_{n}\right\} _{n}$
is explicit due to the first item in the ``furthermore'' part of
Lemma~\ref{lem:LTC-transform}, and has a uniform local corrector with the required running time
due to the second item of that part.\end{myproof}
\begin{remark}
In Lemma~\ref{lem:LTC-transform} above, as in Lemma~\ref{lem:LCC-transform},
we chose to assume that $W$ is $\F_{2}$-linear for simplicity. More
generally, if $W$ is $\F$-linear for any finite field~$\F$, then
$C$ is $\F$-linear as well. Furthermore, the lemma also works if
$W$ is not $\F$-linear for any field~$\F$, in which case $C$
is not guaranteed to be $\F$-linear for any~field~$\F$.
\end{remark}

\subsection{\label{subsec:LTC-transform}Proof of Lemma \ref{lem:LTC-transform}}

Our construction of the LTC~$C$ is the same as the construction
of the LCCs of Section~\ref{subsec:LCC-transform}, with $\tW$ and
$\tau$ replaced by $\dW/2$ and $\delta/2$ respectively. Our LTCs
have the required rate, relative distance and alphabet size
due to the same considerations as before\footnote{In particular,
the lower bound on the relative
distance of our LTC $C$ follows from the lower bound on the relative distance
given in Lemma~\ref{lem:LCC-transform}, using the fact that our
LTC $W$ has a (trivial,
inefficient) $n_W$ query local corrector from $\delta_W/2$ fraction errors.
Again, this lower bound on the distance could have been argued directly,
without talking about locality.}.

It remains to prove that $C$ is locally testable with query complexity~$q\cdot\poly(1/(\varepsilon\cdot\dW))$.
To this end, we describe a local tester~$A$. In what follows, we
use the notation of Section~\ref{sub:LCC-construction}.

Let $\AW$ be the local tester of~$W$. When given oracle access
to a purported codeword $z\in\Sigma^{n}$, the local tester~$A$
emulates the action of~$\AW$ in the natural way: Recall that $\AW$
expects to be given access to a purported codeword of~$W$, and makes
queries to it. Whenever $\AW$ makes a query to a coordinate $j\in\left[\nW\right]$,
the algorithm $A$ performs the following steps:
\begin{enumerate}
\item $A$ finds the block $B_{l}$ to which the coordinate $j$ belongs.
Formally, $l\eqdef\left\lceil j/(b\cdot t)\right\rceil $. 
\item $A$ finds the neighbors of the vertex $u_{l}$ in~$G_{n}$. Let
us denote those vertices by $v_{j_{1}},\ldots,v_{j_{d}}$. 
\item $A$ queries the coordinates $j_{1},\ldots j_{d}$, thus obtaining
the blocks $S_{j_{1}},\ldots,S_{j_{d}}$. 
\item $A$ reconstructs the block $B_{l}$ by reversing the permutation
of $G_{n}$ on $S_{j_{1}},\ldots,S_{j_{d}}$. 
\item If $B_{l}$ is not a codeword of~$\RS_{b,d}$, the local tester~$A$
rejects. 
\item Otherwise, $A$ retrieves the value of the $j$-th coordinate of~$w$
from $B_{l}$, and feeds it to $\AW$ as an answer to its query. 
\end{enumerate}
If $\AW$ finishes running, then $A$ accepts if and only if $\AW$
accepts.

It is easy to see that the query complexity of~$A$ is $d\cdot q$.
It is also not hard to see that if $z$ is a legal codeword of~$C$,
then $A$ accepts with probability~$1$. It remains to show that
if $z$ is not a codeword of~$C$ then $A$ rejects with probability
at least $\dist(z,C)$. To this end, it suffices to prove that $A$
rejects with probability at least~$\frac{1}{\poly(d)}\cdot\dist(z,C)$
-- as explained in Section~\ref{sub:Locally-testable-codes}, this
rejection probability can be amplified to $\dist(z,C)$ while increasing
the query complexity by a factor of $\poly(d)$, which is acceptable.
We use the following definitions: 
\begin{enumerate}
\item Let $S^z_{1},\ldots,S^z_{n}\in\F^{d}$ be the blocks that correspond to
the symbols of $z$. 
\item Let $B^z_{1},\ldots,B^z_{n}\in\F^{d}$ be the blocks that are obtained
from $S^z_{1},\ldots,S^z_{n}$ by reversing the permutation. 
\item Let $w^{z}\in\left(\Lambda\cup\left\{ ?\right\} \right)^{\nW}$ be
the string that is obtained from the blocks $B^z_{1},\ldots,B^z_{n}$
as follows: for each block~$B^z_{l}$ that is a legal codeword of~$\RS_{b,d}$,
we extract from~$B^z_{l}$ the corresponding coordinates of~$w^{z}$
in the natural way. For each block~$B^z_{l}$ that is not a legal codeword
of~$\RS_{b,d}$, we set the corresponding coordinates of~$w^{z}$
to be ``$?$''. 
\end{enumerate}
We would like to lower bound the probability that $A$ rejects~$z$
in terms of the probability that $\AW$ rejects~$w^{z}$. However,
there is a small technical problem: $\AW$ is defined as acting on
strings in $\Lambda^{\nW}$, and not on strings in~$\left(\Lambda\cup\left\{ ?\right\} \right)^{\nW}$.
To deal with this technicality, we define an algorithm~$\AW'$ that,
when given access to a string~$y\in\left(\Lambda\cup\left\{ ?\right\} \right)^{\nW}$,
emulates $\AW$ on $y$, but rejects whenever a query is answered
with~``$?$''. We use the following proposition, whose proof we
defer to Section~\ref{sub:LTC-with-erasures}. 
\begin{proposition}
\label{LTC-with-erasures}$\AW'$ rejects a string~$y\in\left(\Lambda\cup\left\{ ?\right\} \right)^{\nW}$
with probability at least 
\[
\frac{1}{2}\cdot\min\left\{ \dist(y,W),\dW\right\} .
\]

\end{proposition}
\noindent Now, it is not hard to see that when $A$ is invoked on~$z$,
it emulates the action of $\AW'$ on $w^{z}$. To finish the proof,
note that since each coordinate in $W$ has at most $d$ coordinates
of $C$ that depend on it, it holds that 
\[
\dist(z,C)\cdot n\le d\cdot\dist(w^{z},W)\cdot\nW
\]
and therefore 
\begin{eqnarray*}
\dist(w^{z},W) & \geq & \frac{n}{\nW}\cdot\frac{1}{ d}\cdot\dist(z,C)\geq\frac{1}{b\cdot t \cdot d}\cdot\dist(z,C).
\end{eqnarray*}
It thus follows that $A$ rejects~$z$ with probability at least
\[
\frac{1}{2}\cdot\min\left\{ \dist(w^{z},W),\dW\right\} \ge\frac{1}{\poly(d)}\cdot\dist(z,C),
\]
as required.

It is not hard to see that the local tester $A$ can be implemented
efficiently with black box access to~$\AW$, as required by the second
item in the ``furthermore'' part of the lemma.

\subsubsection{\label{sub:LTC-with-erasures}Proof of Proposition~\ref{LTC-with-erasures}}

We use the following result. 
\begin{claim}
\label{LTC-smoothness}Let $I\subseteq\left[\nW\right]$ be a set
of coordinates. The algorithm~$\AW$ queries some coordinate in~$I$
with probability at least 
\[
\min\left\{ \frac{\left|I\right|}{\nW},\frac{1}{2}\cdot\dW\right\} .
\]

\end{claim}
\noindent Note that this claim only makes sense since we assumed that
$\AW$ makes \emph{non-adaptive} queries (we assumed it in Definition~\ref{def:LTC}).
Without this assumption, the probability that $\AW$ queries some
coordinate in~$I$ would have depended on the tested string.
\begin{myproof}
It suffices to prove that for every $I\subseteq\left[\nW\right]$
such that $\frac{\left|I\right|}{\nW}\le\frac{1}{2}\cdot\dW$, the
algorithm~$\AW$ queries some coordinate in~$I$ with probability
at least $\frac{\left|I\right|}{\nW}$. Let $I$ be such a set, and
let $s\in\Lambda^{\nW}$ be an arbitrary string that contains non-zero
values inside $I$, and contains $0$ everywhere outside~$I$. Clearly,
\[
\dist(s,W)=\frac{\left|I\right|}{\nW},
\]
and therefore $\AW$ rejects $s$ with probability at least $\frac{\left|I\right|}{\nW}$.
On the other hand, $\AW$~can only reject~$s$ if it queries some
coordinate in~$I$, since otherwise it cannot distinguish between
$s$ and the all-zeroes codeword. It follows that $\AW$ queries some
coordinate in~$I$ with probability at least $\frac{\left|I\right|}{\nW}$,
as required. 
\end{myproof}
\noindent We turn to proving Proposition~\ref{LTC-with-erasures}.
Let 
\[
E\eqdef\left\{ i:y_{i}=?\right\} 
\]
be the set of erasures in~$y$. We consider two cases: 
\begin{itemize}
\item \textbf{$E$ is ``large'':} Suppose that $\frac{\left|E\right|}{\nW}\ge\frac{1}{2}\cdot\dist(y,W)$.
In this case, it holds by Claim~\ref{LTC-smoothness} that $\AW$
queries some coordinate in~$E$ with probability at least 
\[
\frac{1}{2}\cdot\min\left\{ \dist(y,W),\dW\right\} .
\]
Since $\AW'$ rejects $y$ whenever $\AW$ queries some coordinate
in~$E$, the proposition follows. 
\item \textbf{$E$ is ``small'':} Suppose that $\frac{\left|E\right|}{\nW}\le\frac{1}{2}\cdot\dist(y,W)$.
Let $y_{0}\in\Lambda^{\nW}$ be an arbitrary string that agrees with
$y$ outside~$E$. Clearly, 
\[
\dist(y,W)\le\dist(y_{0},W)+\frac{\left|E\right|}{\nW},
\]
so $\dist(y_{0},W)\ge\frac{1}{2}\cdot\dist(y,W)$. Let $\E$ denote
the event that $\AW$ queries some coordinate in~$E$. We have that
\begin{eqnarray*}
\Pr\left[\AW'\mbox{ rejects }y\right] & = & \Pr\left[\E\right]\cdot\Pr\left[\AW'\mbox{ rejects }y|\E\right]+\Pr\left[\neg\E\right]\cdot\Pr\left[\AW'\mbox{ rejects }y|\neg\E\right]\\
 & = & \Pr\left[\E\right]\cdot1+\Pr\left[\neg\E\right]\cdot\Pr\left[\AW\mbox{ rejects }y_{0}|\neg\E\right]\\
 & \ge & \Pr\left[\E\right]\cdot\Pr\left[\AW\mbox{ rejects }y_{0}|\E\right]+\Pr\left[\neg\E\right]\cdot\Pr\left[\AW\mbox{ rejects }y_{0}|\neg\E\right]\\
 & = & \Pr\left[\AW\mbox{ rejects }y_{0}\right]\\
 & \ge &\dist(y_{0},W)\\
 & \ge & \frac{1}{2}\cdot\dist(y,W),
\end{eqnarray*}
as required. 
\end{itemize}
This concludes the proof.

\subsection{\label{subsec:LTC-sub-poly}Proof of Lemma \ref{high-rate-LTC}}

In this section, we prove Lemma~\ref{high-rate-LTC}, restated below.
\begin{replemma}{\ref{high-rate-LTC}}
There exists an explicit infinite family of
$\F_{2}$-linear codes $\left\{ W_{n}\right\} _{n}$ satisfying: 
\begin{enumerate}
\item $W_{n}$ has block length $n$, rate at least $1-\frac{1}{\log n}$,
and relative distance at least $\exp(-\sqrt{\log n\cdot\log\log n})$.
\item $W_{n}$ is locally testable with query complexity $\exp(\sqrt{\log n\cdot\log\log n})$.
\item The alphabet of $W_{n}$ is a vector space $\Lambda_n$ over $\F_2$, such that 
$|\Lambda_n| \leq \exp\left(\sqrt{\log n\cdot\log\log n}\right)$.
\end{enumerate}
Furthermore, the family $\left\{ W_{n}\right\} _{n}$ has a uniform
local tester that runs in time $\exp(\sqrt{\log n\cdot\log\log n})$.
\end{replemma}

For the proof of Lemma \ref{high-rate-LTC} we use the \emph{tensor product codes} instantiated in the sub-constant relative distance regime. The use of tensor products to construct LTCs was initiated by \cite{BS06}, and was studied further in \cite{V05_tensor_product, DSW06, BV09, BV09b, V11_tensor_robustness}. Our construction is based on a result of \cite{V11_tensor_robustness}.

We start
with some definitions. Let $\F$ be a finite field. For a pair of
vectors $h_{1}\in\F^{\ell_{1}}$ and $h_{2}\in\F^{\ell_{2}}$ their \emph{tensor
product} $h_{1}\otimes h_{2}$ denotes the matrix $M\in\F^{\ell_{1}\times \ell_{2}}$
with entries $M_{(i_{1},i_{2})}=(h_{1})_{i_{1}}\cdot(h_{2})_{i_{2}}$
for every $i_{1}\in[\ell_{1}]$ and $i_{2}\in[\ell_{2}]$. For a pair of
linear codes $H_{1}\subseteq\F^{\ell_{1}}$ and $H_{2}\subseteq\F^{\ell_{2}}$
their \emph{tensor product code} $H_{1}\otimes H_{2}\subseteq\F^{\ell_{1}\times \ell_{2}}$
is defined to be the linear subspace spanned by all matrices of the
form $h_{1}\otimes h_{2}$ where $h_{1}\in H_{1}$ and $h_{2}\in H_{2}$.
For a linear code $H$, let $H^{1}=H$ and $H^{m}=H^{m-1}\otimes H$.
The following are some useful facts regarding tensor product codes
(see e.g. \cite{DSW06}).
\begin{fact}
\label{fact:tensor-codes} Let $H_{1}\subseteq\F^{\ell_{1}}$ and $H_{2}\subseteq\F^{\ell_{2}}$
be linear codes of rates $r_{1},r_{2}$ and relative distances $\delta_{1},\delta_{2}$
respectively. Then $H_{1}\otimes H_{2}\subseteq\F^{\ell_{1}\times \ell_{2}}$
is a linear code of rate $r_{1}\cdot r_{2}$ and relative distance
$\delta_{1}\cdot\delta_{2}$. In particular, if $H\subseteq\F^{\ell}$
is a linear code of rate $r$ and relative distance $\delta$ then
$H^{m}\subseteq\F^{\ell^{m}}$ is a linear code of rate $r^{m}$ and
relative distance $\delta^{m}$. 
\end{fact}
We use the following theorem that is given as Corollary 3.6 in \cite{V11_tensor_robustness}.
\begin{theorem}
[{\label{thm:tensor-test}Immediate corollary of \cite[Thm. 3.1]{V11_tensor_robustness}}]
Let $H\subseteq\F^{\ell}$ be a linear code with relative distance
$\delta$. Then for every $m\geq3$, the code $H^{m}\subseteq\F^{\ell^{m}}$
is locally testable with query complexity 
\[
\ell^{2}\cdot\poly(m)/\delta^{2 m}.
\]

\end{theorem}
For the proof of Lemma \ref{high-rate-LTC}, we instantiate Theorem~\ref{thm:tensor-test}
with the tensor product of Reed-Solomon%
\footnote{We chose Reed-Solomon codes for convenience, but any high-rate codes
with reasonable distance will do.%
} codes. 
\begin{myproof}
[Proof of Lemma \ref{high-rate-LTC}] Fix a codeword length $n\in\mathbb{N}$.
The code $W_{n}$ is defined as follows. Let $\F\eqdef\F_{2^{\sqrt{\log n\cdot\log\log n}}}$,
and let $m\eqdef\sqrt{\frac{\log n}{\log\log n}}$. Let $R$ be a
Reed-Solomon code over $\F$ with block length $n^{1/m}$, rate $r\eqdef\left(1-\frac{1}{\log n}\right)^{1/m}$
and relative distance $1-r$. Note that indeed the block length is
at most $\left|\F\right|$, which is required for the existence of
such codes. Finally, let $W_{n}=R^{m}$.

From the properties of tensor codes we have that $W_{n}$ is a linear
code over $\F$ with block length $(n^{1/m})^{m}=n$, rate $r^{m}=1-\frac{1}{\log n}$,
and relative distance 
\begin{eqnarray*}
\big(1-r\big)^{m} & = & \left(1-\left(1-\frac{1}{\log n}\right)^{1/m}\right)^{m}\\
 & \ge & \left(1-\left(1-\frac{1}{4\cdot m\cdot\log n}\right)\right)^{m} \quad \quad \mbox{(Fact~\ref{power-approximation} : \ensuremath{(1-x)^{y}\le1-\frac{1}{4}\cdot x\cdot y)}}\\
 & = & \left(\frac{1}{4\cdot m\cdot\log n}\right)^{m}\\
 & = & 2^{-O(m\cdot(\log m+\log\log n))}\\
 & = & 2^{-O(\sqrt{\log n\cdot\log\log n})},
\end{eqnarray*}
as required. The fact that $W_{n}$ can be encoded in time $\poly(n)$
follows from standard properties of tensor product codes (see e.g.~\cite[Lecture 6]{S01}).

Finally, by Theorem \ref{thm:tensor-test}, we have that $W_{n}$
is locally testable with query complexity at most 
\[
n^{2/m}\cdot\poly(m) \cdot \left(\frac{1}{4\cdot m\cdot\log n}\right)^{-2 m}=2^{O(\sqrt{\log n\cdot\log\log n})},
\]
as required. The fact that the family $\left\{ W_{n}\right\} _{n}$
has a uniform local tester with the required running time follows immediately from the proof of \cite{V11_tensor_robustness}.
\end{myproof}

\section{Open Questions\label{sec:conclusions}}

We conclude with some open questions.
\begin{itemize}
\item In this work we found that LCCs and LTCs with sub-constant relative distance
can be useful. Are there better LCCs and LTCs in the sub-constant
relative distance regime?
\item LCCs and LTCs often come together with PCPs. Can we construct constant-rate
PCPs with sub-polynomial query complexity?
\item Are there applications of our LCCs and LTCs to complexity theory?\end{itemize}
\begin{acknowledgement*}
We would like to thank Irit Dinur, Tali Kaufman, Ran Raz and Avi Wigderson
for useful discussions and ideas. We would also like to thank Oded
Goldreich, Irit Dinur, Madhu Sudan and anonymous referees for helpful comments
on the preliminary version of this work. 
\end{acknowledgement*}
\bibliographystyle{alpha}
\bibliography{lcc}
 
\end{document}